\documentclass[prb,twocolumn,showpacs,preprintnumbers,amsmath,amssymb]{revtex4}

\usepackage{graphicx}% Include figure files
\usepackage{dcolumn}% Align table columns on decimal point
\usepackage{bm}% bold math

\usepackage{graphics,epsf,psfig}
\usepackage{amssymb,amsmath,amsfonts}
\usepackage{bbm}

\begin{document}

\title{Doping-driven Transition to a Time-Reversal Breaking State 
in the Phase Diagram of the Cuprates}
\author{G. Sangiovanni, M. Capone, S. Caprara, C. Castellani, C. Di 
Castro, and M. Grilli}
\affiliation{Dipartimento di Fisica, Universit\`a di Roma ``La Sapienza'',
and  Istituto Nazionale per la Fisica della Materia (INFM), \\
SMC and Unit\`a  Roma 1, Piazzale Aldo Moro 2, I-00185 Roma, Italy}
\email{massimo.capone@roma1.infn.it}
\date{\today}% It is always \today
%\institute{Dipartimento di Fisica, Universit\`a di Roma ``La Sapienza'',
%and  INFM Center for Statistical Mechanics and Complexity,
%Piazzale Aldo Moro 2, I-00185 Roma, Italy}

\begin{abstract}
Motivated by recent tunnelling and Andreev-reflection experiments, 
we study the conditions for a quantum transition within the 
superconducting phase of the cuprates, 
in which a bulk imaginary (time-reversal breaking) $id_{xy}$ component 
appears in addition to the $d_{x^2 - y^2}$ order parameter. 

We examine in detail the role of some important
 physical features of the cuprates. 
In particular we show that a closed Fermi surface, 
a bilayer splitting, an orthorhombic distortion, 
and the proximity to a quantum critical point around optimal doping 
favor the appearance of the imaginary component.
These findings could explain why the mixed $d_{x^2 - y^2}+ id_{xy}$
order parameter is  observed in YBCO and LSCO, and suggest that it
could appear also in Bi2212.
We also predict that, in all cuprates, the mixed state
should be stable only in a limited doping region all contained beneath 
the  $d_{x^2 - y^2}$ dome. The behavior of the specific heat at
 the secondary transition is discussed.
\end{abstract}

\maketitle

\section{Introduction}

The evidence for a $d_{x^2 - y^2}$ symmetry 
of the superconducting pairing in the High-T$_c$ Cuprate 
Superconductors (HTCS) is nowadays overwhelming\cite{tsueirev}.
Nonetheless it is well known that a secondary gap component of different 
symmetry can develop either spontaneously or driven by external factors, 
like a magnetic field, magnetic impurities as well as 
the proximity to an interface \cite{laughlin,balatsky1,buchh,matsumoto}.
The first important consequence is that the nodal lines of the gap
located along the diagonals 
of the Brillouin zone ($\Gamma\mbox{X}$ direction) can be 
removed, deeply altering the low energy properties. Secondly,
 this kind of superconducting state can break the 
time-reversal symmetry with important consequences 
on the physical properties\cite{sigrist}. 
Specifically a superconducting state with mixed $d_{x^2 - y^2}+id_{xy}$ 
symmetry has been invoked in Refs. \cite{laughlin,balatsky1,euroghosh}
to explain the thermal conductivity anomalies in 
$\mbox{Bi}_2\mbox{Sr}_2\mbox{Ca}\mbox{Cu}_2\mbox{O}_8$ (Bi2212) \cite{krishana,movshov},
in Ref. \cite{franz} to account for the
properties of single quasiparticle states observed in the vortex 
core \cite{fischer}, and in Ref. \cite{vojta} as a 
possible explanation of the momentum and temperature dependence of 
the damping of quasiparticles. Scanning tunnelling 
experiments in $\mbox{Y}_{1-y}\mbox{Ca}_y\mbox{Ba}_2\mbox{Cu}_3\mbox{O}_{7-x}$
 (YBCO) \cite{tunnelexp,deutscher} and in  
$\mbox{La}_{2-x}\mbox{Sr}_x\mbox{Cu}\mbox{O}_4$ (LSCO) \cite{gonnelli} 
have also shown clear deviations from the pure $d_{x^2 - y^2}$-wave 
symmetry. While the LSCO results have been originally interpreted 
in terms of a $d+s$ wave, a recent reanalysis based on the results of the 
present report has shown that the
$d_{x^2 - y^2}+id_{xy}$ symmetry is a more faithful description 
of the experiments\cite{gonnelli2}.

Our main motivation comes indeed from tunnelling experiment on  
YBCO films \cite{deutscher}, which  have evidenced that the 
zero-bias conductance peak associated with a pure 
$d_{x^2 - y^2}$ gap function\cite{hukashiwaya} undergoes a 
{\it doping-} and {\it magnetic-field-dependent} splitting.
This can be interpreted \cite{fogeltanuma} in terms of 
a secondary time-reversal breaking component of the superconducting gap.
It is quite important to notice that 
the amplitude of the splitting has a power-law dependence on  
doping  and magnetic-field. The (zero-field) splitting is 
in fact linear in the deviation from the optimal doping for overdoped samples,
and it is zero for underdoped samples. Moreover,
starting from a finite (zero) value at zero field, the splitting
increases linearly in the overdoped (underdoped) materials with
 the magnetic field, while it displays a square-root dependence
close to optimal doping. These power-law dependencies are
strong indications of a critical behavior due to a second-order
phase transition and are hardly accounted for by
surface or disorder effects alone. 
The above result suggests therefore a doping-driven 
{\it bulk} transition.
The observed fraction of $d_{xy}$ pairing increases with 
$\delta$, but it must be noticed that it never exceeds the $20\%$ of 
the principal one in all the measured samples\cite{deutscher}. 
These experimental situation leaves the door open for at least two
scenarios. In the first one, discussed in a different context 
 in Ref. \cite{sachdev},
the $d_{xy}$ component keeps increasing by further doping and eventually
the gap becomes of pure $d_{xy}$ symmetry.
In the  alternative scenario, which is supported by our analysis,
the $d_{xy}$ fraction does not monotonically increase, 
but it starts to decrease at some
doping, leading to a smaller dome beneath the $d_{x^2 - y^2}$,
where the mixed order parameter is stable.
Our scenario is also consistent with other experiments which
put severe upper bounds for any imaginary component of the gap 
(see Ref. \cite{tsueirev} for a recent discussion).

The experimental framework described above suggested us to carry out
a systematic analysis of 
the specific properties which may favor the occurrence of a mixed 
superconducting state in some cuprates and not in others. 
To achieve this goal, we quantitatively  analyze the role of
band-structure effects (Van Hove singularity (VHS), bilayer 
splitting, and orthorhombic distortion), and  
of the form of the pairing interaction in determining the weight of the 
secondary component (if any), 
the features of the phase diagram and the behavior of observables like the
specific heat.   
We rely on the simple BCS approach to make the role of the
above ingredients more transparent. We notice that the fact that 
the $d_{x^2 - y^2}+id_{xy}$ order parameter, when observed \cite{deutscher},
establishes in the overdoped regime, where strong correlation effects are
less important.

In Sec. II we introduce the formalism and discuss in different subsections the effects of the Fermi surface (A), the bilayer splitting (B), the orthorhombic
distortion (C) and of a QCP (D). In Sec. III we discuss how a phase diagram 
for the secondary component can be drawn, and Sec. IV is devoted to the
conclusions.

\section{Method and Results}

General theoretical constraints for the appearance of a bulk  mixed order parameter
have already been studied, e.g., in Refs. \cite{matsumoto,ghosh}.  
In particular, in the first of these works it has been
shown that the mixed state is more likely to establish close to a 
surface than in the bulk and that the $id_{xy}$ wave is the most favored 
secondary component in both cases.

Here we recall some basic symmetry arguments and derive the BCS equations.
The strong anisotropy of the HTCS allows us to focus on a 
purely two-dimensional lattice system, characterized by the ${\cal{C}}_{4v}$
point group.
Only the four harmonics corresponding to the one-dimensional irreducible 
representations of the group are compatible with singlet pairing.
We can write them as\cite{ostlund} 
\begin{eqnarray*}      
w^{s^+}_{{\bf r};{\bf k}}&=&\cos{k_x x}\cos{k_y y} + \cos{k_x y}\cos{k_y x},\\
w^{s^-}_{{\bf r};{\bf k}}&=&-\sin{k_x x}\sin{k_y y}+ \sin{k_x y}\sin{k_y x},\\
w^{d^+}_{{\bf r};{\bf k}}&=&\cos{k_x x}\cos{k_y y} - \cos{k_x y}\cos{k_y x},\\
w^{d^-}_{{\bf r};{\bf k}}&=&-\sin{k_x x}\sin{k_y y}- \sin{k_x y}\sin{k_y x}.
\end{eqnarray*}
Then, the gap function can be written as
\begin{equation} \label{deltaeq}
\Delta_{\bf k} = \sum_{\eta} \Delta_{\bf k}^\eta =
\sum_{\eta} {\sum_{\bf r}}^\prime \Delta_{\bf r}^\eta
w_{{\bf r};{\bf k}}^\eta,
\end{equation}
where the index $\eta = s^+, s^-, d^+, d^-$ labels the different 
representations and ${\bf r}= (x,y)$, with integer $x,y$, denotes a 
lattice site (we take the lattice spacing $a_x=a_y=1$).
The primed sum is restricted to inequivalent sites under the symmetry of the
lattice (i.e., to the different lattice distances). 
If we require the invariance of $|\Delta_{\bf k}|^2$
under the point group transformations,
the gap function has to transform either like a single representation, 
or like a complex (time-reversal breaking) combination of the form
$\Delta^\eta_{\bf k} + i \Delta^\zeta_{\bf k}$ with $\eta \neq \zeta$.

Expanding the BCS equation in the same harmonics, it is easy
to realize that
the appearance of a harmonic labeled by 
${\bf r}$ is controlled by the correspondent component of the 
pair potential $V({\bf r})$.
In a generic model for the cuprates the on-site interaction ($V_0$) 
is repulsive (positive) due to strong local repulsion. 
The longer distance part ($V_{i \neq 0}$) is instead attractive (negative) and
it is reasonably assumed to be a decreasing function of distance.
We can therefore restrict to the smallest distance harmonics.
In the following we always use the symbol $V_i$ to denote the
modulus of the $V(\bf r = \bf i)$.  
For ${\bf r} = 0$, the only nonzero harmonic belongs to the  $s^+$
representation, and 
corresponds to the isotropic $s$-wave, which is ruled out by the
repulsive $V_0$. For ${\bf r}=(\pm 1,0);(0, \pm 1)
\equiv {\bf 1}$ the 
$s^+$ and $d^+$ representations are associated with the extended 
$s_{x^2+y^2}$ and $d_{x^2 - y^2}$ waves, respectively. For 
${\bf r}= (\pm 1,\pm 1)\equiv {\bf 2}$, 
the non-vanishing harmonics are related to the 
$s_{xy}$ and $d_{xy}$ waves ($s^+$ and $d^-$ representations). 

If we take a complex combination of the form 
$\Delta_1 w^{\eta}_{{\bf 1};{\bf k}}+i\Delta_2 w^{\zeta}_{{\bf 2};{\bf k}}$, 
with both $\Delta_1$ and $\Delta_2$ real,
each gap parameter affects the other only through the quasiparticle spectrum 
and the secondary pairing can be viewed as a BCS coupling between the 
Bogoljubov quasiparticles of the primary superconducting phase.
The BCS equations then read
\begin{eqnarray}
{1\over V_1}&=& 
\int {d{\bf k}\over 4\pi^2}
\left(w^{\eta}_{{\bf 1};{\bf k}}\right)^2 {\tanh (\beta E_{\bf k}/2)\over 2 E_{\bf k}}\equiv I_1\label{BCS1}\\
{1\over V_2}&=& 
\int {d{\bf k}\over 4\pi^2}
\left(w^{\zeta}_{{\bf 2};{\bf k}}\right)^2 {\tanh (\beta E_{\bf k}/2)\over 2 E_{\bf k}}\equiv I_2
\label{BCS2}
\end{eqnarray}
where $\beta$ is the inverse temperature,
$E_{\bf k}=[\xi_{\bf k}^2+(\Delta_1 w^{\eta}_{{\bf 1};{\bf k}})^2+
(\Delta_2 w^{\zeta}_{{\bf 2};{\bf k}})^2]^{1/2}$, with $\eta=s^+,d^+$, 
$\zeta=s^+,d^-$ ($\eta \neq \zeta$).
The band dispersion is $\xi_{\bf k}=-2t(\cos k_x+\cos k_y)+4t^\prime\cos k_x \cos k_y-\mu$ where $t$ and $t'$ are nearest and next-nearest neighbor 
hoppings and $\mu$ is the chemical potential. 
Density functional theory gives a typical value for the hopping $t$ 
in the HTCS of  $200$ m$eV$, while $t^\prime$ assumes different values 
in the various compounds\cite{andersen}. 
These values can however be reduced by a factor of two by correlation effects,
that can be estimated by fitting the angular resolved photoemission spectra
with a tight-binding model. 
As we will see later in more detail, the value of  $t^\prime$ has important 
effect on the DOS, since it controls the position of the Van Hove singularity 
(VHS).

Being the largest attractive coupling $V_1$ controls the principal 
component of the gap, which can be $d_{x^2 - y^2}$ or $s_{x^2 + y^2}$. 
For typical values of the particle density in the HTCS, 
both low doping and the VHS of the two-dimensional 
DOS strongly favor the $d_{x^2 - y^2}$ symmetry, 
which thus represents the principal component of the superconducting gap 
even in the simplified BCS approach \cite{marki}. 
We take henceforth $\eta=d^+$ in Eqs. (\ref{BCS1}),(\ref{BCS2}).
The spectrum for the secondary $\zeta$ component is therefore 
(pseudo)gapped by the principal $d_{x^2-y^2}$ gap $\Delta_1$. 
As a consequence, the mixed order parameter establishes only
for $V_2$ larger than a critical value $V_2^{cr}$, as opposed to the 
case of the Cooper instability in the metallic phase, which 
takes place for arbitrarily small coupling. 
The critical coupling for both $d_{xy}$ and $s_{xy}$ secondary pairing 
is determined by  solving the BCS equations (\ref{BCS1}),(\ref{BCS2}) 
with $\Delta_2=0$ and $\zeta=d^-,s^+$ respectively. 
The only difference between these two cases is the different form of the 
harmonic $w^{\zeta}_{{\bf 2};{\bf k}}$ that weights the same kernel in $I_2$.
This kernel, in turn, is affected by the presence of the  $d_{x^2-y^2}$ 
wave. The $d_{x^2-y^2}$ gap in the $E_{\bf k}$ spectrum strongly suppresses 
the contributions from the ${\mbox M}$ points $(0,\pm \pi),(\pm\pi,0)$,
while it does not affect much the nodal lines. 
The $d_{xy}$ wave 
has nodes along the $\Gamma{\mbox M}$ directions and is maximum along the 
diagonals, while the $s_{xy}$ has its maximum contributions from the 
${\mbox M}$ points, just like the $d_{x^2-y^2}$. 
It is therefore clear that the $d_{xy}$ is 
most likely candidate for the secondary component \cite{matsumoto}.

A numerical solution of the above Eqs. (\ref{BCS1}),(\ref{BCS2}) 
confirms the above arguments, finding $V_2^{cr}$ always larger for 
$s_{xy}$ than for $d_{xy}$. 
Moreover, the appearance of the $s_{xy}$ component is generally
associated to a first-order transition from pure $d_{x^2-y^2}$ to 
pure $s_{xy}$, since the continuous mixing is unlikely due to the
strong competition between the two harmonics. 
This kind of scenario is clearly incompatible with the experiments,
in which the secondary component appears in a continuous way. 
The continuous transition from $d_{x^2 - y^2}$ to $d_{x^2 - y^2}+id_{xy}$ 
gap is therefore the most natural candidate if a pure $d_{x^2 - y^2}$ is to be 
modified by a small secondary component, as suggested by the 
experiments on YBCO\cite{deutscher} and on LSCO \cite{gonnelli}.

Next, we evaluate the critical value  of $V_2$ to get $d_{xy}$ pairing, 
$V_{d_{xy}}^{cr}\equiv V^{cr}$, assuming a given value of $V_1$.
A similar analysis has been performed in Ref. \cite{matsumoto}
for a continuous model, and in Ref.\cite{ghosh} for a lattice model.
In the latter case, in which the effect of a two-dimensional DOS is
considered, it is always found that $V^{cr} > V_1$.
This would imply that a secondary 
component is possible only taking a potential which increases with 
the distance, at least going from nearest to next-nearest 
neighbors. This condition seems hard to be realized
in a microscopic model, where the interaction is naturally 
a decreasing function of distance.
Similar results have been obtained also within a $t$-$J_1$-$J_2$ model in 
\cite{sachdev}, where the  $d_{x^2-y^2}+id_{xy}$ and pure 
$d_{xy}$ establish, respectively, for $J_2/J_1  \simeq 1.4$ and $2.2$,
which are clearly not representative of the cuprates.
In the case of Ref. \cite{matsumoto}, $V^{cr}/V_1$ may be smaller than one
for some values of the parameters, but this only comes from the
assumption of a flat DOS, which is clearly not representative of
the HTCS.

Having in mind that for a plausible microscopic potential  
$V_2 < V_1$, in the following we explore the conditions for the
appearance of the secondary harmonic obeying this constraint.
First we notice that the larger is the value of $V_1$ is, the
smaller is the ratio $V^{cr}/V_1$. In other words,
if both the couplings are large, a secondary component may appear also
for $V_2 < V_1$.
The above claim can be understood as follows:
If we increase $V_1$, $\Delta_1$ also increases, and
it has different effects on the two equations.
The integral  $I_2$ in Eq. (\ref{BCS2}) does not change much by increasing
$\Delta_1$, due to the $w^{d^-}_{{\bf 2};{\bf k}}$ factor,
while $I_1$ instead contains $w^{d^+}_{{\bf 1};{\bf k}}$, and it is 
therefore substantially reduced by $\Delta_1$.
Since, from Eqs. (\ref{BCS1}),(\ref{BCS2}), $V^{cr}/V_1 = I_1/I_2$, 
this implies that, at fixed band dispersion, the larger is $V_1$, the smaller
is the ratio $V^{cr}/V_1$ and then the mixed state is 
more favored.

The argument above shows that strong coupling is necessary for 
the appearance of a secondary component\cite{notaghosh}. Nevertheless, 
$V_1$ can not be increased indefinitely 
without pushing the value of $\Delta_1$ to unphysical values.
If we want to increase the $d_{x^2-y^2}$-coupling at fixed $\Delta_1$ we can 
introduce a cut-off $\omega_0$ in the integrals.
This cut-off is not just an artifact to push the system to stronger 
coupling, but naturally measures the typical energy scale
of the attractive potential. In the forthcoming calculations we take
$\omega_0 = 50$ m$eV$ $= 0.25t$.
\cite{notacutoff}. 
As we see in the following, this choice of cut-off leads us to 
reasonable phase diagrams without assuming unphysical values of the
gap. 

Now we discuss the role of different physical effects
on the value of the critical $V_2$.
Some preliminary considerations may be done before numerically
solving the BCS equations.
In a BCS approach, the bandstructure influences the values of the gap and of
the critical temperature only through the DOS. 
In particular, the VHS plays the main role, strongly favoring the pure
$d_{x^2-y^2}$ pairing. 
Other harmonics may be stabilized only for doping values for which the
chemical potential is sufficiently far from the VHS. 
This means that, if we want the
$d_{xy}$ to appear for dopings, like the ones of the
experiments, the VHS must be relatively close to half-filling.
This is realized for cuprates with closed Fermi surface like LSCO.
To explain the appearance of the secondary component in YBCO 
we therefore have to invoke some other effects, like orthorhombic
distortion or bilayer splitting.
Both this effects are expected to help the secondary component
to appear, because they ``regularize'' the VHS.
We discuss the role of these effects in the following subsections.
The final subsection is instead dedicated to the form of the
pairing potential $V(\bf{r})$, and in particular to the possible
role of a QCP in the charge sector close to optimal doping\cite{notaqcp}.

\subsection{Shape of the Fermi Surface}
\begin{figure}
\begin{center}
\includegraphics[width=9cm]{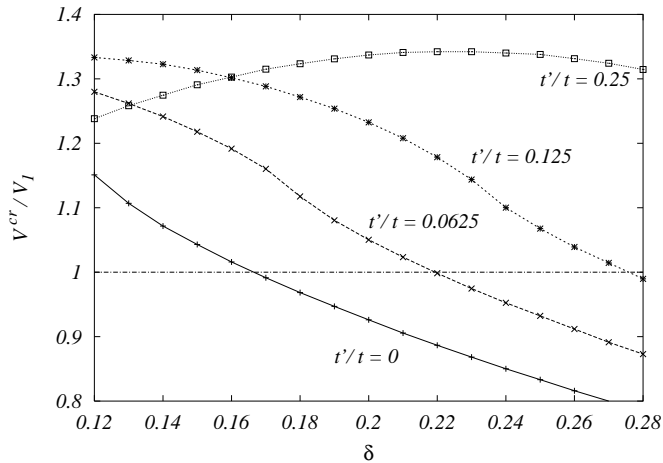}
\caption{$V^{cr} / V_1$ as a function of the doping $\delta$ for 
$V_1/t = 1.15$, $\omega_0/t=0.25$  and different values of the ratio
$t^\prime/t$. When the curve crosses the line $V^{cr}/V_1=1$ 
the appearance of the secondary harmonics becomes possible with
$V_2 < V_1$.}
\label{tprimo}
\end{center}
\end{figure}
As we mentioned above, the position of the VHS plays an important role
in determining the symmetry of the order parameter within the BCS approach.
The ratio $t'/t$ is a direct measure of the position of the VHS, and 
consequently of the shape of the Fermi surface. For $t'=0$, the
VHS lies at the chemical potential at half-filling.
Increasing $t'/t$, the density for which the chemical potential lies 
at the VHS moves away from half-filling.
Consequently, the stability of the pure $d_{x^2-y^2}$ wave extends to larger
doping, making the appearance of the $d_{xy}$ more difficult.
We numerically solve Eqs. (\ref{BCS1}) and (\ref{BCS2}) to make
the analysis more quantitative. The results are reported in Fig. 1.
Here we take $V_1$ constant as 
a function of doping, just to avoid the complications of many
varying parameters and to disentangle the effect of the VHS.
It is understood that $V_1$ will depend on doping in a realistic
description of the cuprates.

According to the discussion above ($V_2 < V_1$) , 
the secondary harmonics can only appear, roughly speaking, when
$V^{cr}/V_1 < 1$.
The appearance of the secondary harmonic is more likely as the 
doping increases, but the  minimal doping 
$\delta_{min}$ above which $V^{cr}/V_1 < 1$ increases with 
$t^\prime/t$.
$\delta_{min}$ assumes  values compatible with the 
experimental findings  only for (small) values of $t^\prime/t$ such that 
the Fermi surface is closed, while  an 
open Fermi surface pushes $\delta_{min}$ to exceedingly large values. 
This analysis suggests that the 
$d_{x^2 - y^2}+id_{xy}$ pairing may likely occur in LSCO, where the Fermi 
surface is expected to be closed,
at least in the optimal and overdoped region.
Other physical effects must be invoked for open Fermi surface materials
like YBCO or Bi2212.

\subsection{Bilayer Splitting}
Materials like  YBCO and Bi2212 have relatively large values 
$t^\prime/t \gtrsim 0.25$, and therefore  open Fermi surfaces, 
which favor, according to the previous analysis, a pure $d_{x^2-y^2}$
pairing.
However, the strict two-dimensionality assumed up to now
overestimates the effect of the VHS, which strongly favors the 
$d_{x^2 - y^2}$ wave.
YBCO and Bi2212, as opposed to LSCO, are indeed multi-layer cuprates.
Introducing the multi-layer structure of Bi2212 and YBCO results in 
a ``regularization'' of the VHS. 
ARPES measurements in Bi2212 show indeed a bilayer splitting 
$2t_\perp$ at the {\mbox M} points which, although smaller than the one 
predicted by band-structure calculations \cite{andersen}, is sizable, and 
varies from 88 m$eV$ to 140 m$eV$ \cite{bilayer}. We are not aware of 
similar data on YBCO (where ARPES measurements are much harder to
perform), but the bare values of $t_\perp$ in Bi2212 and YBCO 
are similar, so that we can expect similar renormalized values.

\begin{figure}
\begin{center}
\includegraphics[width=9cm]{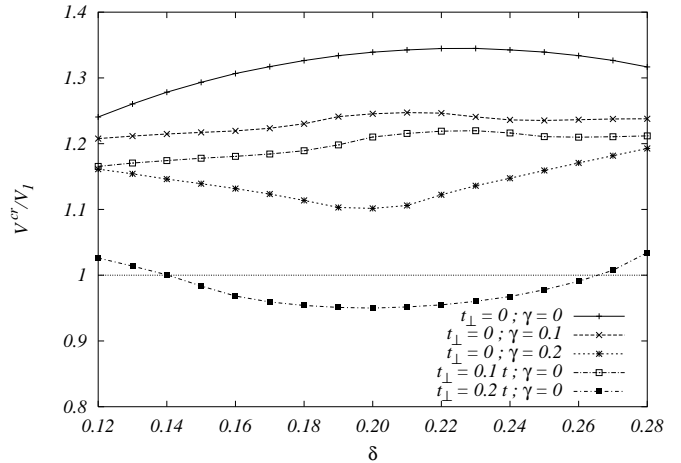}
\caption{$V^{cr} / V_1$ as a function of the doping $\delta$ for 
$V_1/t = 1.15$, $\omega_0/t=0.25$  and different values of the splitting
parameter $t_\perp$ and of the orthorhombic parameter $\gamma=(t_y-t_x)/t_x$. 
The upper curve is the same as that shown in Fig.\ref{tprimo}.}
\label{nuovafig}
\end{center}
\end{figure}

To analyse the role of a bilayer structure we introduce a band splitting 
of the form ${1\over 2}t_\perp (\cos k_x-\cos k_y )^2$ which modifies 
the effective DOS and splits the VHS partially
spoiling the $d_{x^2-y^2}$ principal 
component. On the other hand, the nodal regions are not 
affected and so one expects the ratio 
$V^{cr} / V_1$ to decrease as soon as $t_\perp \neq 0$.
In Fig.\ref{nuovafig} we show the doping dependence of $V^{cr} / V_1$ for the 
same values of $V_1$ and of $\omega_0$ as in the case of Fig.\ref{tprimo}.
The two curves with open and solid squares corresponds to the case
$t_\perp = 0.1 t$ and $0.2 t$ respectively and are, as expected, lower 
than that for $t_\perp = 0$.  
As can be seen in the figure, the second of these curves, which still 
compatible with the measured value of $t_{\perp}$ is Bi2212, has been lowered
enough to cross the $V^{cr} / V_1 = 1$ line. 

In conclusion, the bilayer splitting helps
the $d_{xy}$ component to establish at least in all those doping regions for 
which the chemical potential falls in the dip of the density of states centered
 at $-4t'$, leaving one of the two Fermi surfaces open. This is a relevant 
scenario for the case of YBCO and Bi2212 in which the Fermi surfaces is open.

\subsection{Orthorhombic Distortion}

Another important effect that lowers the impact of the VHS on the 
pairing in the cuprates is an orthorhombic splitting of the 
bands. This is accounted for by taking a dispersion of the form 
$\xi_{\bf k}=-2 (t_x \cos k_x+ t_y \cos k_y)+
4t^\prime\cos k_x \cos k_y-\mu$ and $t_x \ne t_y$.
In this way $\xi_{\bf k}$ and the kernel containing
$E_{\bf k}=[\xi_{\bf k}^2+|\Delta_{\bf k}|^2]^{1/2}$ that enters the equations 
(\ref{BCS1}),(\ref{BCS2}), are no longer symmetric functions with
respect to the ${\cal{C}}_{4v}$ group. 
In any case this kernel can still be expanded  
in terms of the ${\cal{C}}_{4v}$-harmonics and one can solve the resulting 
coupled equations.
However, if the orthorhombic distortion is small, one can 
neglect the effects on the symmetry of the gap and take only the larger
effects
on the density of states. In doing this, one neglects the $d^+$-$s^+$ 
and $d^-$-$s^-$ mixing that arises because the ${\cal{C}}_{4v}$-group  
symmetry of the system has been lowered to ${\cal{C}}_{2v}$. 
In fact, the basis functions for  the two one-dimensional representations 
of ${\cal{C}}_{2v}$ that couples to the singlet are given by 
$w^{s^{\pm}}_{{\bf r};{\bf k}} + w^{d^{\pm}}_{{\bf r};{\bf k}}$ respectively,
and what we do in the following is to project our solutions on the original 
$d_{x^2-y^2}$ and $d_{xy}$ harmonics in order to understand the role of a 
small orthorhombic distortion on the $d_{x^2-y^2} + id_{xy}$ solution.

The density of states is then qualitatively similar to the bilayered one. 
In fact the VHS is split and one expects that when the chemical 
potential lies in the dip the secondary $d_{xy}$ component is again favored
with respect to the case with $t_x=t_y$. 
The results are represented by the two curves with crosses and stars in 
Fig.\ref{nuovafig}. We have used for $t_x$ the same value of $200$ m$eV$ 
as before for $t$ and varied the ratio $(t_y-t_x)/t_x$ denoted as $\gamma$. 
$V_1$ and $\omega_0$ are still those of the Fig.\ref{tprimo}. 
In the case of $\gamma=0.1$ (crosses) the ratio
$V^{cr} / V_1$, although lower than for the isotropic one, is still greater 
than $1$, as well as that for $\gamma=0.2$ (stars) even if in this last one
the characteristic upward curvature coming from the presence of 
the dip in the DOS can be clearly seen.
As for the bilayer case we have used a constant value for $V_1$ aiming 
at analysing the effects of band structure of the orthorhombic distorted 
compounds on the critical coupling, and we have evidenced that the $d_{xy}$
secondary component is favored by this effect. 
The values of $\gamma$ that we have used are representative of the case of 
the cuprates according to, e.g., 
Ref.\cite{sorella} where a value of $\gamma \sim 0.1$ is found by 
means of {\it ab-initio} calculations for the case of LSCO.

\subsection{Quantum Criticality around Optimal doping}

We now discuss how the form of the pair potential can be important
to obtain the  $d_{x^2-y^2}+id_{xy}$ order parameter.
In particular, we discuss how this  phenomenology can be
related to the QCP scenarios for HTCS\cite{scenari1,prl95,scenariafm}. 
In this context, it has been suggested that the pairing may be 
mediated by quasi-critical 
fluctuations close to the QCP in the charge and/or spin sector(s)
located near optimal doping.
In this framework the correlation function of the critical fluctuations 
determines the functional form of the pairing potential.
A few different QCP have been proposed, ranging from magnetic and/or
charge ordering, to time reversal breaking. In the last part of this section
we briefly discuss the case of charge and spin ordering, 
while here we first discuss some
general consequences of a QCP near optimal doping which are independent
on the nature of the ordered phase.  
Far from criticality (i.e., far from optimal doping)
the short-range components $V_1,V_2,\ldots$ may 
decrease strongly  with the distance. 
At criticality the correlation saturates below a non-universal length scale, 
and all the components of the potential below this scale become
of the same order of magnitude. 
$V_2$ then becomes of the order of $V_1$, favoring the appearance of a 
secondary component of the gap. 
This establishes the 
connection between the $d_{x^2-y^2} \to d_{x^2-y^2}+id_{xy}$
transition and the underlying quantum criticality.
One may notice that, close to the QCP, where various couplings are 
of the same order of magnitude, higher-order harmonics should 
in principle be considered within each irreducible 
representation. The main outcome of this extension 
would only be a more detailed description of the gap profile, 
without changing  the symmetry of the order parameter. 
Observed deviations from a pure $d_{x^2-y^2}$ gap can be explained 
in this context \cite{pera}.

We now consider the specific case of a pairing interaction mediated
by the fluctuations around a charge ordering QCP, with an 
incommensurate ordering vector ${\bf Q} = (Q_x,Q_y)$. 
The effective interaction between quasiparticles in the Cooper channel
can be written as
\begin{equation} \label{potcr}
V_{{\bf k}-{\bf k'}}= U -\frac{1}{8}\sum_{\alpha}\frac{V_c}{\kappa^2+\omega_{{\bf k}{\bf k'}}({\bf Q}^\alpha)},
\end{equation}
where $U$ is a residual local repulsion, $V_c$ is the attraction
strength, $\kappa^2$ is a ``mass term'' measuring  the distance from 
criticality, and  $\omega_{{\bf k}{\bf k'}}({\bf Q}^\alpha) = 2(2-\cos 
(k_x-k'_x-Q_{x}^\alpha)-\cos (k_y-k'_y-Q_{y}^\alpha))$ contains the 
momentum dependence of the interaction. The sum over the eight momenta
${\bf Q}^\alpha=(\pm Q_{x},\pm Q_{y}),(\pm Q_{y},\pm Q_{x})$ makes the
interaction symmetric under the ${\cal{C}}_{4v}$ group.

The real-space expression for (\ref{potcr}) is particularly useful in
light of the previous analysis. We can in fact write
\begin{equation} \label{realspace}
V({\bf r}) = U \delta_{{\bf r},0}-\frac{1}{2} V_{c} \alpha_{\bf r}(\kappa^2) 
w^{s^+}_{{\bf r};{\bf Q}}. 
\end{equation}
The local repulsion obviously enter only in the on-site
coupling and does not therefore play any role in the 
$d_{x^2-y^2} \to d_{x^2-y^2}+id_{xy}$ transition.
$\alpha_{\bf r}(\kappa^2)$ are dimensionless functions of $\kappa^2$ 
and the lattice distances {\bf r}. 
When the system is at the critical point, i.e., in  the limit $\kappa^2 \to 0$,
all the $\alpha_{\bf r}$ approach the same value regardless of the index 
{\bf r}, as we expected from the general arguments given above. 
The dependence on the critical vector {\bf Q} is entirely contained in the 
symmetry factor $w^{s^+}_{{\bf r};{\bf Q}}$, i.e., in the above defined 
harmonics computed at the ordering vector {\bf Q}. 
The first two coupling constants that enter the BCS equations 
(\ref{BCS1}) and (\ref{BCS2}) are then given by 
$V_1=\frac{1}{2} V_c \alpha_{\bf 1} w^{s^+}_{{\bf 1};{\bf Q}} = 
\frac{1}{2} V_c \alpha_{\bf 1} (\cos Q_x + \cos Q_y)$ and 
$V_2=\frac{1}{2} V_c \alpha_{\bf 2} w^{s^+}_{{\bf 2};{\bf Q}} =
V_c \alpha_{\bf 2} \cos Q_x \cos Q_y$ 
\cite{notaalpha}.
The modulation of the couplings introduced by the critical momentum
{\bf Q} determines regions in which some coupling may become repulsive,
thus making the correspondent harmonic impossible to establish.
In particular, the $d_{x^2-y^2}+id_{xy}$ symmetry is possible only if both
$V_1$ and $V_2$ are attractive, i.e., for  $Q_x \in [-\pi/2:\pi/2]$ and $Q_y 
\in [-\pi/2:\pi/2]$. 
Moreover, the ratio $V_2/V_1$ is proportional to the form factor
$2\cos Q_x \cos Q_y/(\cos Q_x + \cos Q_y)$, and it is therefore maximum
at ${\bf Q} = (0,0)$. The mixed symmetry order parameter is therefore
more likely for small ordering vectors.

The contribution to the pairing interaction coming from the spin sector can 
be written in the same way as (\ref{potcr}), with an opposite sign.
This implies that pairing mechanisms based on antiferromagnetic
spin fluctuations\cite{scenariafm}, 
with ordering vector ${\bf Q}=(\pi/2,\pi/2)$
strongly favor $d_{x^2-y^2}$ pairing, which is maximum for this vector,
while the $V_2$ coupling is repulsive for the same vector.
Even taking into account small incommensurations around the antiferromagnetic
vector, spin fluctuations mechanisms do not favor
the appearance of the $d_{x^2-y^2}+id_{xy}$ order parameter.
Because of the relevance of antiferromagnetic spin fluctuations in the 
underdoped region of the phase diagram of the cuprates this analysis 
indicates that the mixed $d_{x^2-y^2}+id_{xy}$ state should not be present 
at least in the extremely underdoped compounds, in agreement with 
the experiments.

To summarize the main outcome of this section,
the proximity to a QCP helps in meeting the necessary conditions for
the mixed symmetry (i.e., strong coupling and $V_2/V_1$ close to 1).
Nevertheless, the specific nature of the QCP (like, e.g., 
charge and/or spin ordering) may introduce sign modulations of the 
real-space interaction, making  some components repulsive,  
and therefore  suppressing the correspondent harmonics. 
The pairing mechanism based on antiferromagnetic spin 
fluctuations \cite{scenariafm} strongly favors the  
$d_{x^2-y^2}$ pairing, but suppresses the $d_{xy}$ channel,
leading to pure $d_{x^2-y^2}$.
Mechanisms based on charge fluctuations support instead 
a secondary $d_{xy}$ component besides a primary $d_{x^2-y^2}$
provided the characteristic ordering
wave vector $\bf{Q}$ is not too large ($|{\bf Q}| < \pi/2$).

\section{Phase Diagram}

\begin{figure}
\begin{center}
\includegraphics[width=9cm]{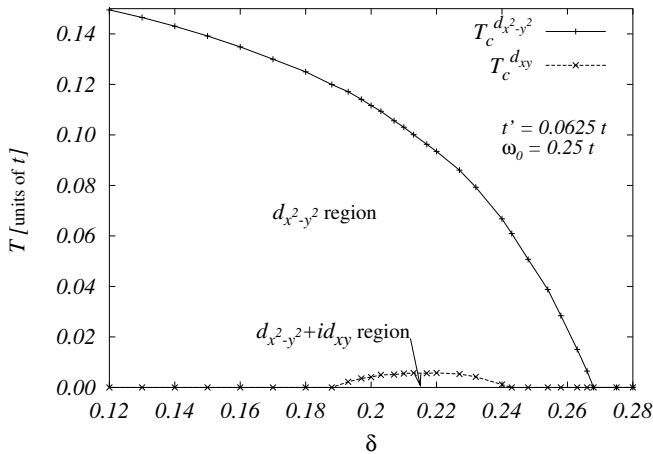}
\caption{Phase diagram for $d_{x^2 - y^2}$ and $d_{x^2-y^2} + id_{xy}$
pairing in the HTCS. The parameters, which are
appropriate for open Fermi surface cuprates, are given in the text.}
\label{phd}
\end{center}
\end{figure}
We can now draw a phase diagram for the $d_{x^2-y^2} + id_{xy}$ pairing 
by fixing $V_1$ for each doping $\delta$, such that the mean field 
critical temperature for $d_{x^2-y^2}$ pairing (which we interpret as the
temperature for Cooper pair formation without phase coherence in the 
underdoped regime) grossly follows the behavior of the pseudogap temperature 
$T^*$ for small values of $\delta$, and becomes negligibly small
for the doping at which $T_c$ vanishes in the overdoped regime. 
In this way we phenomenologically introduce  the
doping dependence of the principal component of the interaction 
$V_1(\delta)$. In particular $V_1(\delta)$ will
be larger at low doping and decrease toward weak coupling in the
overdoped regime. Around optimal doping we obtain $V_1/t \simeq 2 $, which is 
larger than the value we used to draw the diagrams of Figs. 1 and 2,
and, according to the previous discussion on the relevance
of the strong coupling, is more favorable for the 
$d_{x^2 - y^2}+ id_{xy}$ pairing.

On the other hand, we consider the less favorable 
situation in terms of bandstructure by using the single-layer isotropic 
dispersion ($t_\perp=0$ and $\gamma=0$). 
According to the previous analysis, in order to have a chance to observe
a $d_{xy}$ component in this unfavorable situation, we need to assume a small
value of $t'=0.0625t$ which gives a closed Fermi surface in the relevant 
doping range. 
The value of the interaction is obviously crucial. 
Here we take, at any doping,  $V_2/V_1 = 0.8$, a relatively large value, which, as we have 
seen, can be reasonably reached if a QCP is located close to optimal doping.
This constant value of $V_2/V_1$ allows us to analyse the combined role of 
the density of states and of the doping dependence of the principal coupling. 
In Fig. \ref{phd}, we show $T_c$ for $d_{x^2-y^2}$ pairing
(top curve) described above, and the critical temperature for the 
onset of the secondary pairing (bottom curve), calculated by solving
the coupled BCS equations (\ref{BCS1}) and (\ref{BCS2}).
As shown in Fig. \ref{phd}, the mixed $d_{x^2 - y^2}+id_{xy}$ order 
parameter is stabilized in a small dome slightly shifted in
the overdoped region, $0.19 < \delta < 0.24$, 
and the $d_{xy}$ gap is always only a small fraction of the 
$d_{x^2-y^2}$ gap.

The shape of the mixed state region is the consequence of the 
balance between two main effects: {\it {i)}} the $d_{xy}$ can 
establish only when the VHS is far enough from the chemical 
potential, and therefore appears only above a given value of the
doping. {\it {ii)}} Further overdoping, $V_1(\delta)$ 
decreases and pushes the system toward weak coupling, where
the mixed order parameter ceases to be stable for reasonable
values of $V_2/V_1$.
The dome shape is therefore an unavoidable consequence of
taking a sensible value for the ratio $V_2/V_1 < 1$.
Only by releasing this physical condition, and taking $V_2/V_1 \gg 1$ the
$d_{xy}$ component may indefinitely increase by overdoping, 
eventually leading to a pure $d_{xy}$ pairing
in the heavily overdoped region \cite{sachdev}. 
The re-analysis of the LSCO data\cite{gonnelli} with a $d_{x^2-y^2} +id_{xy}$ 
gap parameter, clearly displays the closure of the dome in the overdoped 
region, in qualitative agreement with our analysis. 

We note that small variations of the ratio $V_2/V_1$ can significantly
modify the amplitude of the secondary component.
At doping $\delta=0.22$, for example, by slightly enhancing $V_2/V_1$ 
from 0.8 to  0.9, the critical 
temperature for the $d_{x^2-y^2} + id_{xy}$ state is enhanced by a factor of 
$3$.

Up to now, we have found that the $d_{x^2-y^2} \rightarrow
d_{x^2-y^2} + id_{xy}$ transition can be found for values of the parameters
appropriate to the single layer isotropic case with open Fermi surface.
The discussions of Sec. II B, C have shown that, as soon as small
perturbations of the pure two-dimensional bandstructure are considered, 
the constraints for the onset of the secondary component are
partially released. The data of Fig. 2 show in fact that for 
a realistic value of the bilayer splitting, the whole region
$0.14 < \delta < 0.26$ becomes compatible with a $d_{x^2-y^2} +id_{xy}$ 
order parameter. Analogously, a small orthorhombic distortion 
lowers the values of the critical coupling.
Once realistic bandstructure parameters involving the above
effects are used, the appearance of the time-reversal breaking 
order parameter in LSCO and YBCO, as well as the doping dependence, 
can be easily explained in the simple BCS approach.
The physical mechanisms leading to the mixed order parameter here presented 
have however a general content which does not depend explicitly on the 
use of the BCS approach.
It is important to stress that the 'dome behavior' of the 
mixed order parameter does not depend on the details of the calculations, 
and represents a generic result within our approach.

The experimental studies which motivated our work identified the 
transition to the mixed $d_{x^2-y^2}+id_{xy}$ wave with the 
splitting on the zero bias conductance peak.
Here we finally discuss the possibility to provide a further experimental 
signature of the mixed order parameter by measuring the variation of 
the specific heat coefficient $\gamma_{c_V} = c_V/T$
at $d_{x^2-y^2} \to d_{x^2-y^2}+id_{xy}$ transition.
For the parameters used to draw the phase diagram of Fig. 3, we have
computed the mean-field jump of the specific heat at $\delta =0.22$, where the
secondary component is maximum, and the effect of the second transition
should be more evident. 
We actually computed the entropy as a function of temperature, and
extracted the specific heat from $c_V= T dS/dT$.

In this specific case, at the transition from the metal to the 
$d_{x^2-y^2}$ superconductor we obtain  $\Delta\gamma_{c_V}/\gamma_{c_V} = 
1.00$. 
At the $d_{x^2-y^2}+id_{xy}$ transition (which occurs approx. at $10$ K 
in this case) we obtain instead $\Delta\gamma_{c_V}/\gamma_{c_V} = 0.28$.
We note that the absolute value of $\Delta\gamma_{c_V}$ in the latter
case is about two orders of magnitude smaller then in the 
former, higher temperature transition.
Even though fluctuation effects are expected to round the mean-field jump, 
our estimated value of  $\Delta\gamma_{c_V}$
suggests that the effect could be still experimentally detected.

\section{Conclusions}
We have performed a systematic analysis of the possibility of a 
$d_{x^2-y^2} +id_{xy}$ symmetry pairing in the cuprates by means of
the simple BCS approach which is indeed justified by the
experimental evidence that the mixed order parameter
establishes in the overdoped region. 
Our results are compatible with all the available indications
of $d_{x^2 - y^2}+ id_{xy}$ pairing in YBCO and LSCO.
We have separately discussed the role of the bandstructure effects and
of the pairing potential. 
Since the VHS is an obstacle to the mixed pairing, all the effects
that lower the impact of the singularity result in an enhanced 
tendency toward the $d_{x^2-y^2} +id_{xy}$ pairing.
In particular, the bilayer splitting characteristic of multilayer
cuprates and the orthorhombic splitting of the bands both favor the
mixed order parameter.
The relevance of the bilayer splitting suggests that,
Bi2212 is also likely to show a mixed order parameter, with a small
mixing of the same order of YBCO.

As far as the pairing potential is concerned, we have shown that 
a mixed-symmetry superconducting state is more likely
to establish at strong coupling. In particular,
it can appear for $V_2/V_1 < 1$ only at strong coupling.
The presence of a QCP  
around optimal doping is an important ingredient, since it 
enhances the  coupling  and makes the
next-nearest neighbor component of the pairing potential 
(responsible for the $d_{xy}$ pairing) of the same order of the
nearest neighbor one (responsible for $d_{x^2-y^2}$).

The main result of our analysis is that a
 mixed order parameter of the form $d_{x^2-y^2} +id_{xy}$ can
be stabilized in a small dome contained in the larger $d_{x^2-y^2}$ region
with a  pairing potential being a decreasing function of the distance, i.e., 
$V_2 < V_1$.
Contrary to other scenarios, the $d_{xy}$ component does not grow 
indefinitely with increasing doping
and it is always smaller than the $d_{x^2-y^2}$.  
The only way to obtain an indefinitely increasing $d_{xy}$ component 
is to take an unphysically large ratio $V_2/V_1 \gg 1$.

We have also shown that $d_{x^2-y^2} +id_{xy}$ pairing
is more likely to occur for materials with closed Fermi surfaces
such as LSCO, and in compounds where an 
inter-layer hopping splits the VHS.
The less important role of the VHS in the electron-doped materials 
may lead to the  $d_{x^2-y^2} +id_{xy}$ state also in these
family of compounds.
We have also discussed the impact of the secondary transition on 
specific heat measurements, showing that the relative jump of the
linear specific heat coefficient is a fraction of the jump
at the primary (higher temperature) transition to the $d_{x^2-y^2}$
superconducting phase.

\begin{center}
***
\end{center}

We thank D. Daghero, R.S. Gonnelli and G.A. Ummarino for useful 
discussions and for having shown us the reanalysis of their data prior 
to publication.
Discussions with A. Toschi are also acknowledged.
We acknowledge financial support by
MIUR Cofin 2001, prot. 2001023848, and by INFM/G (through PA-G0-4).


\begin{thebibliography}{999}
\bibitem{notaghosh} In light of this discussion, we can understand 
the large values of the critical secondary coupling 
($V^{cr} > V_1$)  
obtained in Ref. \cite{ghosh} as due to the weak-coupling regime
in which all the calculations are done.
\bibitem{notaqcp} We emphasize here that the QCP we are referring to is 
responsible for the fluctuations mediating pairing, and it is different
from the possible $d_{x^2-y^2} \to d_{x^2-y^2}+id_{xy}$ QCP.
In particular it does not (necessarily) involve time-reversal 
symmetry breaking.
\end{thebibliography}
\end{document}